\input epsf
\documentstyle[preprint,eqsecnum,floats,aps]{revtex}

\begin{document}

\title{\hfill FERMILAB-Pub-97/402\\
Calibrating  the energy of a 50 $\times$ 50 GeV muon
collider using spin precession}

\author{ Rajendran Raja\\
             $\&$\\
         Alvin Tollestrup\\
       {\it Fermi National Accelerator Laboratory } \\
       {\it P.O. Box 500 } \\
       {\it Batavia, IL 60510 }}

\maketitle

\begin{abstract}
The neutral Higgs boson is expected to have a mass in the region 
90-150 GeV/c$^2$ in various schemes within the Minimal Supersymmetric
extension to the Standard Model. A first generation Muon Collider is uniquely
suited to investigate the mass, width and decay modes of the Higgs boson, since
the coupling of the Higgs to muons is expected to be strong enough for it to be
produced in the $s$ channel mode in the muon collider. Due to the narrow width
of the Higgs, it is necessary to measure and control the energy of the
individual muon bunches to a precision of a few parts in a million.
We investigate the feasibility of determining the energy scale of a muon
collider ring with circulating  muon beams of 50 GeV energy by measuring the
turn by turn variation of the  energy deposited by electrons produced by the
decay of the muons. This variation is caused by the existence of an average
initial polarization of the muon beam and a non-zero value of $g-2$ for the
muon. We demonstrate that it is feasible to determine the energy scale of the
machine with  this method to a few parts per million using data collected
during  1000 turns.
\end{abstract}
\section{The method}
The spin vector $\vec{S}$ of a muon in the muon collider will precess according
to the following equation, first derived by Bargmann, Michel and Telegdi
\cite{bargman}
\begin{equation}
%\begin{array}{l}
 \frac{d\vec{S}}{dt} = \vec{\Omega} \times \vec{S} \\
%\end{array}
\end{equation}
\begin{equation}
 \vec{\Omega} = - \frac{e}{\gamma m_\mu}\left( (1+ a\gamma)\vec{B}_\bot
+ (1 + a)\vec{B}_\| - (a\gamma + \frac{\gamma}{1+\gamma})\vec{\beta}\times
\frac{\vec{E}}{c} \right)
\label{bmt}
\end{equation}
where $\vec{B}_\bot$ and $\vec{B}_\|$ are the transverse and parallel 
components of the magnetic field with respect to the muon's velocity
$\vec{\beta}c$, $e$ is the electric charge, $m_\mu$ the  mass of the muon, $a
\equiv \frac{g-2}{2} $ is the magnetic moment anomaly of the muon and  $\gamma$
and $g$ are the Lorentz factor and the gyromagnetic ratio of the muon. The value
of $a \equiv \frac{g-2}{2} $ for the muon is 1.165924E-3  \cite{pdg}. 
In what follows, we will consider the ideal planar collider ring case 
where $\vec{B}_\|$ and $\vec{E}$ are zero. For such a collider ring,
$\vec{\Omega}$ is given by 
\begin{equation}
 \vec{\Omega} = \vec{\Omega}_{cyc}(1+a\gamma)
\end{equation}
where $\vec{\Omega}_{cyc}$ is the angular velocity of the circulating beam. From
this, it follows that when the beam completes one turn, the spin will rotate 
by a further $a\gamma \times 2\pi$ radians. We will compute the precision with
which $\gamma$ can be determined by measuring the energy of the electrons
produced by muon decay in this ideal case.
We will examine the  effects of departures from the ideal case in the 
last section.

 It can be shown that the angular distribution of the decay electrons
in the muon center of mass is given by the relation \cite{barr}
\begin{equation}
 \frac {d^2N}{dx dcos\theta} = N(x^2(3-2x) - \hat{P} x^2(1-2x) cos\theta)
\label{eq1}
\end{equation}
where $N$ denotes the number of muon decays, $x \equiv 2E/m_\mu $  is the
electron energy $E$ in the muon rest frame expressed as a fraction  of the
maximum possible energy ($\approx 0.5m_\mu$), $cos\theta$ is the angle of the
electron in the muon rest frame with respect to the $z$ axis which is the
direction of motion  of the muon in the laboratory and $\hat P$ is the product
of the muon  charge and the $z$ component of the muon polarization. The muon
polarization is defined as the average of the individual muon unit spin vectors
over the ensemble of muons considered. We note that the distribution is linear
in $\hat P$.

\begin{figure}[p]
\epsfxsize = 16.cm
\epsffile{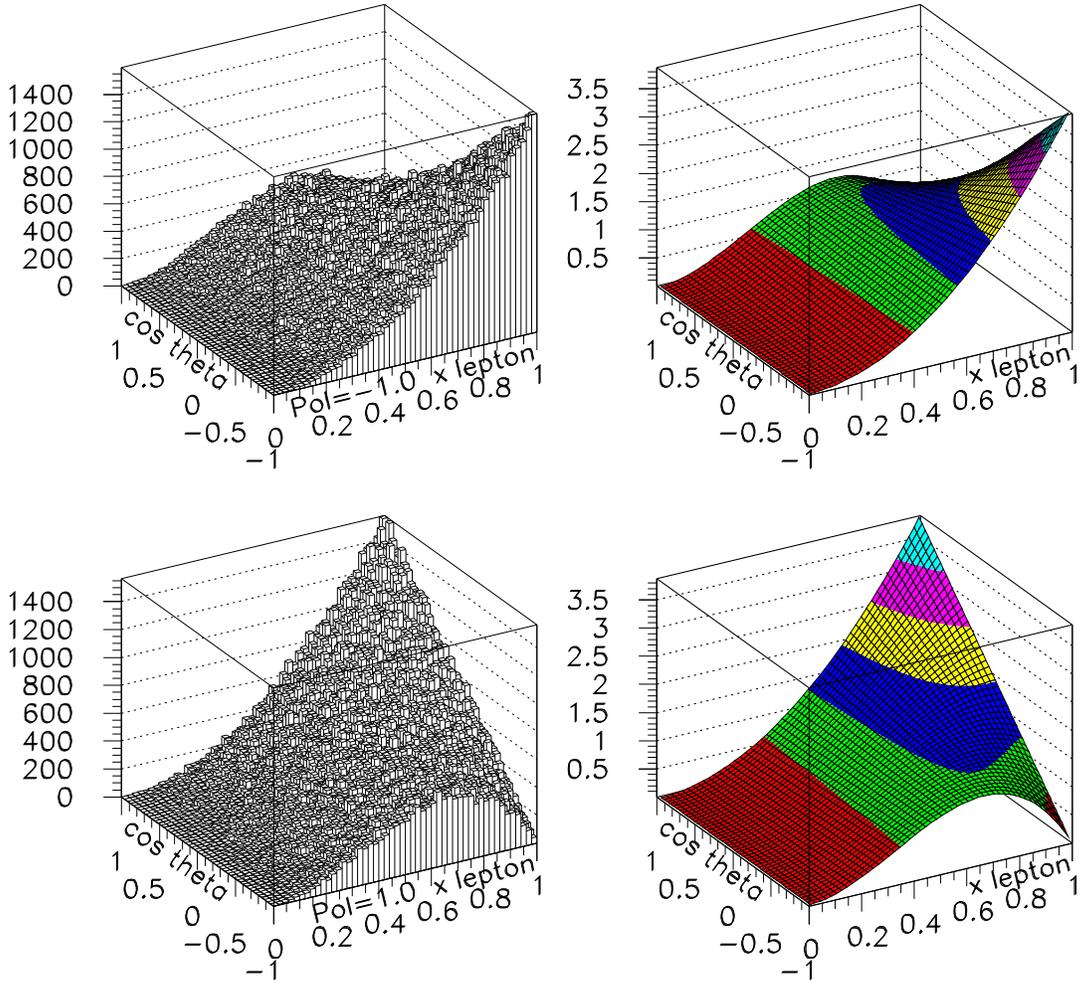}
\caption{The top lego plots shows the generated events and the theoretical decay
function in the $x,cos\theta$ plane for ${\hat P}$ = -1.0. The lego plots at the
bottom of the figure show the corresponding plots for ${\hat P}$ = 1.0. }
\label{fig1}
\end{figure}

 A routine was written to generate muon decays according to equation \ref{eq1}.
Figure \ref{fig1} shows the shape of the function in equation \ref{eq1} and the
generated events in $x,cos\theta$ space for various values of $\hat P$. 
There is excellent agreement between the theoretical shape of the
function and the Monte Carlo generated events.
\begin{figure}[p]
\epsfxsize = 16.cm
\epsffile{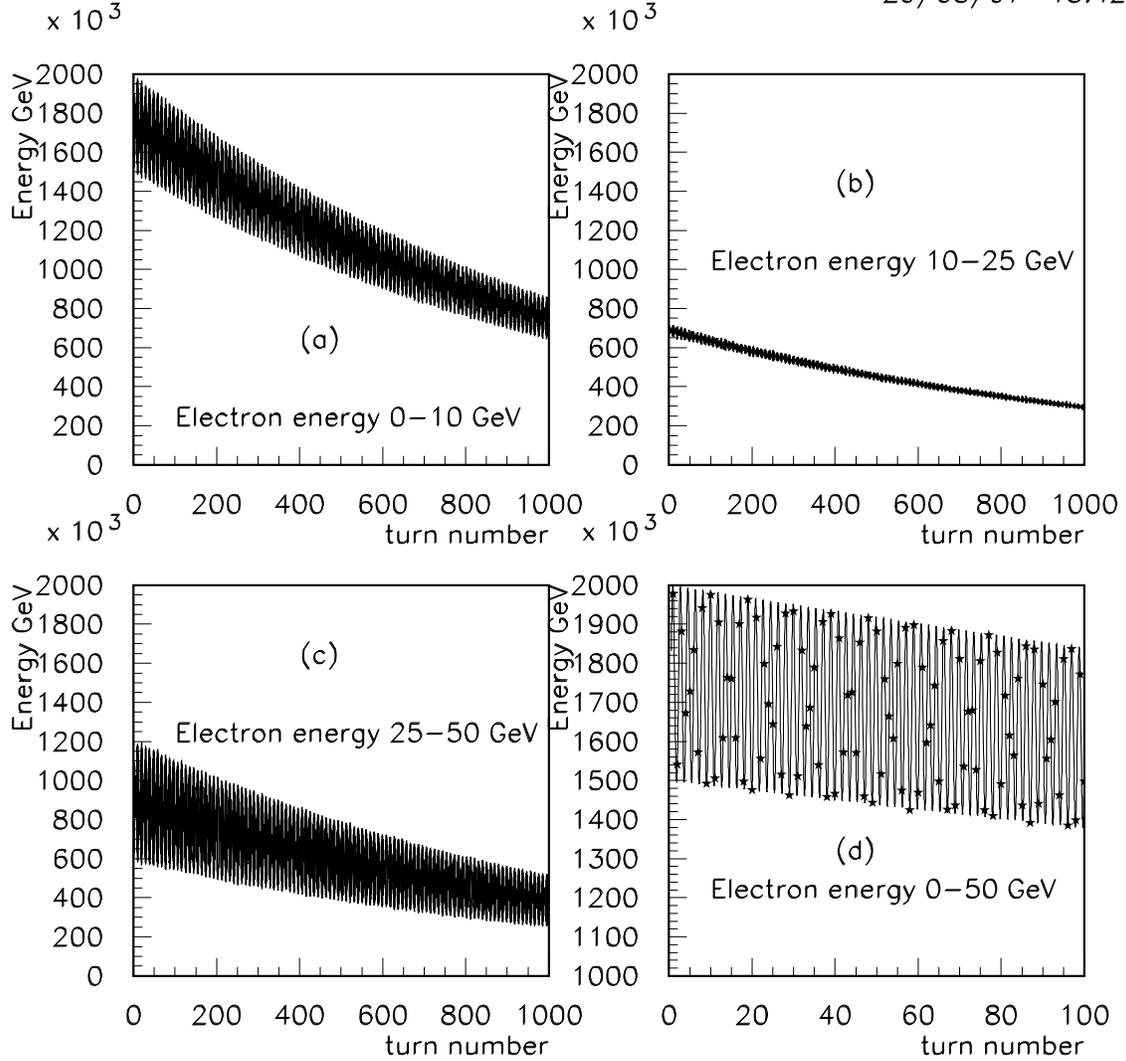}
\caption{(a)Total energy observed as a function of turn number for ${\hat P}$ =
-1.0 with  individual electron energies in the range 0-10 GeV for 100,000 muon
decays. (b) Electron energies in the range 10-25 GeV (c) 25-50 GeV (d) All
electrons included. Superimposed is a functional form defined by equation
\ref{eq2}}
\label{fig2}
\end{figure}
The average energy $<E>$ and longitudinal momentum $<P_L>$ of the electron in
the muon rest frame can be obtained using equation \ref{eq1} as follows.
\begin{eqnarray}
 <E> = \frac{m_\mu}{2} \int  \int x \frac {d^2N}{dx dcos\theta} dx dcos\theta
  = \frac{7}{10} \frac{m_\mu}{2} \\
 <P_L> = \frac{m_\mu}{2}\int\int x cos\theta
 \frac {d^2N}{dx dcos\theta} dx dcos\theta
 = \frac{\hat P}{10} \frac{m_\mu}{2}
\end{eqnarray}
 These two quantities form the components of a 4-vector, whose transverse
 components are zero, which may be
 transformed to the laboratory frame to yield the average electron energy
 $<E_{lab}>$.
\begin{equation}
 <E_{lab}> = \frac{7}{20}E_\mu (1+ \frac{\beta}{7} {\hat P})
\end{equation}
 where $E_\mu$ is the energy of the muon beam .
 Since the polarization $\hat P$ precesses from turn to turn by the amount
 $\omega = \gamma (g-2)/2 \times 2\pi $ radians, and the number of muons
 decrease turn by turn due to decay and losses, the total energy $E(t)$ due
 to decay electrons  observed during turn $t$   in an
 electromagnetic calorimeter will have the following expression
\begin{equation}
 E(t) = N e^{(-\alpha t)}(\frac{7}{20}E_\mu (1+ \frac{\beta}{7}
 ({\hat P}cos \omega t+ \phi)))
\label{eq2}
\end{equation}
where $N$ is the number of muon decays sampled in turn 0, $\phi$ is an 
arbitrary phase containing information on the initial direction of polarization
and  $\alpha$ is the turn by turn decay constant of the
muon intensity which in the absence of losses other than decay is given by
\begin{equation}
 \alpha = \frac{t_{circ}}{\gamma t_{life}} 
\end{equation}
where $t_{circ}$ is the time taken to circulate around the storage ring and
$t_{life}$ is the muon life time. 

 For a 100\% polarized beam, the amplitude of the oscillations is only $1/7$
that of the non-oscillating background.
It can be seen from equation \ref{eq1} that the sensitivity to ${\hat P}$ is
enhanced by selecting larger values of $cos\theta$. This implies selecting
electrons with higher laboratory energy. 
Figures \ref{fig2}(a-c) show the deposited electron energy as a function of turn
number for  polarization ${\hat P} = 1.0$ for individual electron energy ranges
of 0-10 GeV, 10-25 GeV and 25-50 GeV respectively as a function of turn number.
Figure 2(b) shows very little  oscillatory signal, since the electrons in that
energy range have small values of $cos \theta$. Figure 2(d) shows the deposited
electron energy with no electron energy cuts. Superimposed is the predicted
behavior according equation \ref{eq2}. This serves as a consistency check for
our routines.  The signal to background ratio increases as we demand electrons
with higher value of $cos\theta$. In what follows, we use electrons with 
energy greater than 25 GeV during the investigative phase of this analysis
and will later optimize this cut. In practice, we can select electrons with
energies above a value by momentum analyzing them with a dipole field
before they enter the calorimeter.

 The method to determine the energy scale of the collider would then entail
fitting a functional form of the type
\begin{equation}
  f(t) = A e^{-B t}(C cos(D + E t) + F)
\label{eq3}
\end{equation}
 to the energy observed in the calorimeter. The variables $A,B,C,D,E,F$ are
parameters to be fitted. The information on the energy scale is contained in the
parameter $E$. 
\subsection{Parameters of a 50 GeV idealized muon storage ring}
In order to arrive at reasonable numbers for $\alpha$ and $\omega$, we consider
a storage ring of 50 GeV muons with a uniform bending field of 4 Tesla. This
would produce a circular ring with the parameters given in table \ref{tab1}.
\begin{table}[t]
\begin{center}
{\footnotesize
\begin{tabular}{|c|c|c|c|}
\hline
 Parameter & Value & Parameter & Value \\
\hline
Muon Energy & 50 GeV &
$\gamma$     &     473.22 \\
spin precession in one turn &    3.4667 radians &
Magnetic field & 4.0 Tesla\\
radius of ring &        41.66666 meters &
beam circulation time &     0.87327E-06 sec \\
dilated muon life time &    0.10397E-02 sec &
turn by turn decay constant &  0.8399E-03 \\
\hline
\end{tabular}
}
\end{center}
\caption{ Parameters of an idealized muon storage ring}
\label{tab1}
\end{table}
It should be noted that for an idealized storage ring with constant B field 
considered here, $\alpha$ does not depend on $\gamma$, since
\begin{eqnarray}
 t_{circ} = \frac{m_\mu\gamma}{0.3 B c}\\
 \alpha = \frac{2\pi m_\mu}{0.3 B c t_{life}}
\end{eqnarray}
where $m_\mu$ is the muon rest mass, B is the bending field of the storage ring
and $c$ is the velocity of light. A 100 GeV collider ring will have the same
$\alpha$ as a 50 GeV collider ring or a 25 GeV collider ring in this idealized
case. As $\gamma$ changes slightly, $t_{circ}$ changes in proportion, $\alpha$
being the constant used to convert measurements of $t_{circ}$ to $\gamma$.
Measuring the decay rate of muons also affords a second method to determine
$\gamma$.
The beam circulation time $t_{circ}$  can be measured to precisions of the order
of a part in $10^6$ and  the fractional error in muon lifetime is 1.82E-5
\cite{pdg}. The fractional error in $\gamma$ obtainable by observing  the rate
of  decay of the muons will then be dominated by the precision that one can
measure $\alpha$, namely $\delta\gamma/\gamma = \delta\alpha/\alpha$. 

\subsection{Generation of events and fitting for $\gamma$}
Since  equation \ref{eq1} is linear in $\hat P$, the decay distribution of an
ensemble of muons depends only on $\hat P$, the ensemble average of the $z$
component of  the individual muon spin vectors.  However, because of the
momentum spread of the muons, each individual particle will have a $\gamma$
slightly different from the average and hence the precession of the spin vector
around the ring will be  different, leading to a slightly different
value of $\hat P$ for the next turn.  We model the beam by generating an
ensemble of 100,000 muons each having its own spin vector and momentum.  In an
actual collider, it will be possible to sample significantly more decays than
this.  During each turn, we decay all the beam particles  once and record the
number and total energy deposited by electrons with individual energies above 25
GeV. Approximately 27$\%$ of the decay electrons pass this cut, on average. We
decrease the number of decays by the appropriate number expected by muon decay
alone for the next turn. At this stage we do not introduce fluctuations in the
number of decays from turn to turn, since the 100,000 muons are meant to be
representative of  a much larger number in the actual ring. 
We precess the 100,000 spin vectors by their individual precession rates and
make them decay again. We repeat this for 1000 turns. We re-use the muons after
each turn since the 100,000 muons represent our model of the muon ensemble in
the collider.
\subsubsection{Generation of muon spin vectors} 
\begin{figure}[p]
\epsfxsize = 16.cm
\epsffile{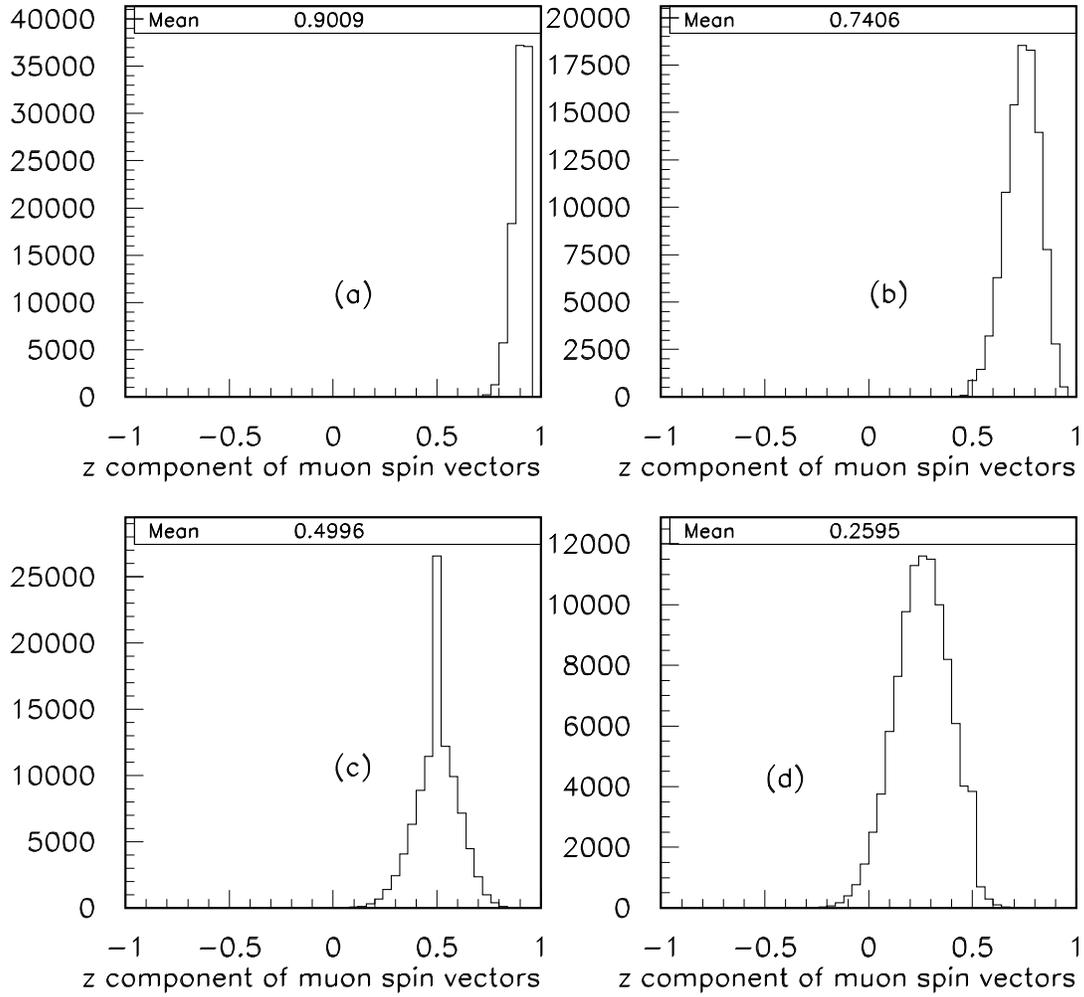}
\caption{(a)-(d) show the distribution of the z component of the spin
vectors for  the four samples considered.}
\label{fig0}
\end{figure}

We generate 4 different  samples of events with different ensembles of spin
vectors. The z component of the unit spin vector of a muon $S_z$ is allowed to
vary from -1 to 1. This range is divided into 51 bins and  the z components are
generated using a binomial distribution whose average value is specified.  We
are justified in treating this problem in this classical fashion, since each
``muon" represents an ensemble of actual muons with quantized spin components. A
more realistic generation of the spin vectors with correlations between momentum
spread and $\hat P$ would require a detailed modelling of the pion decay and
muon transport systems and is not warranted here since the effect due to the
distribution in $S_z$ is expected to be small. Figures  \ref{fig0} (a-d) show
the distributions of $S_z$ for the 4 samples. The average value of the
distributions are 0.9, 0.74, 0.5 and 0.26 respectively. We study negatively
charged muons resulting an initial value of $\hat P$ of -0.9,-0.74,-0.5 and
-0.26  respectively for these samples. In the absence of momentum spread, the
decay distributions would only depend on $\hat P$ and not on the details of the
distribution of $S_z$. The  angles of the spin vectors are precessed by the 
individual $\gamma$ dependent precession rate from turn to turn. In what 
follows, we assume a beam energy spread of 0.03$\%$ for the muons for all 
samples unless otherwise specified.

\subsubsection{Fitting procedure and generation of errors}
The energy deposited every turn is fitted to the functional form given by
equation  \ref{eq3} using the CERN program MINUIT \cite{minuit}. In order to
study the variation of the fractional error $\delta\gamma/\gamma$ with the
number of electrons sampled, we fluctuate the energy observed in the
calorimeter $E_m$ by 

\begin{equation}
\frac{\sigma^2_{E_m}}{<E_m>^2} \approx \frac{1}{N}(1.03153 )
\end{equation}
where $N$ is the number of electrons sampled. See Appendix for a derivation of
this formula. We analyze the case for 41261, 10315, 2579 and 1146 electrons
sampled which corresponds to a fractional error in the measured total energy of
PERR $ \equiv \frac{\sigma_{E_m}}{E_m}$ of .5E-2,1.0E-2,2.0E-2 and 3.0E-2
respectively.
\section{Results}
\begin{figure}[p]
\epsfxsize = 16.cm
\epsffile{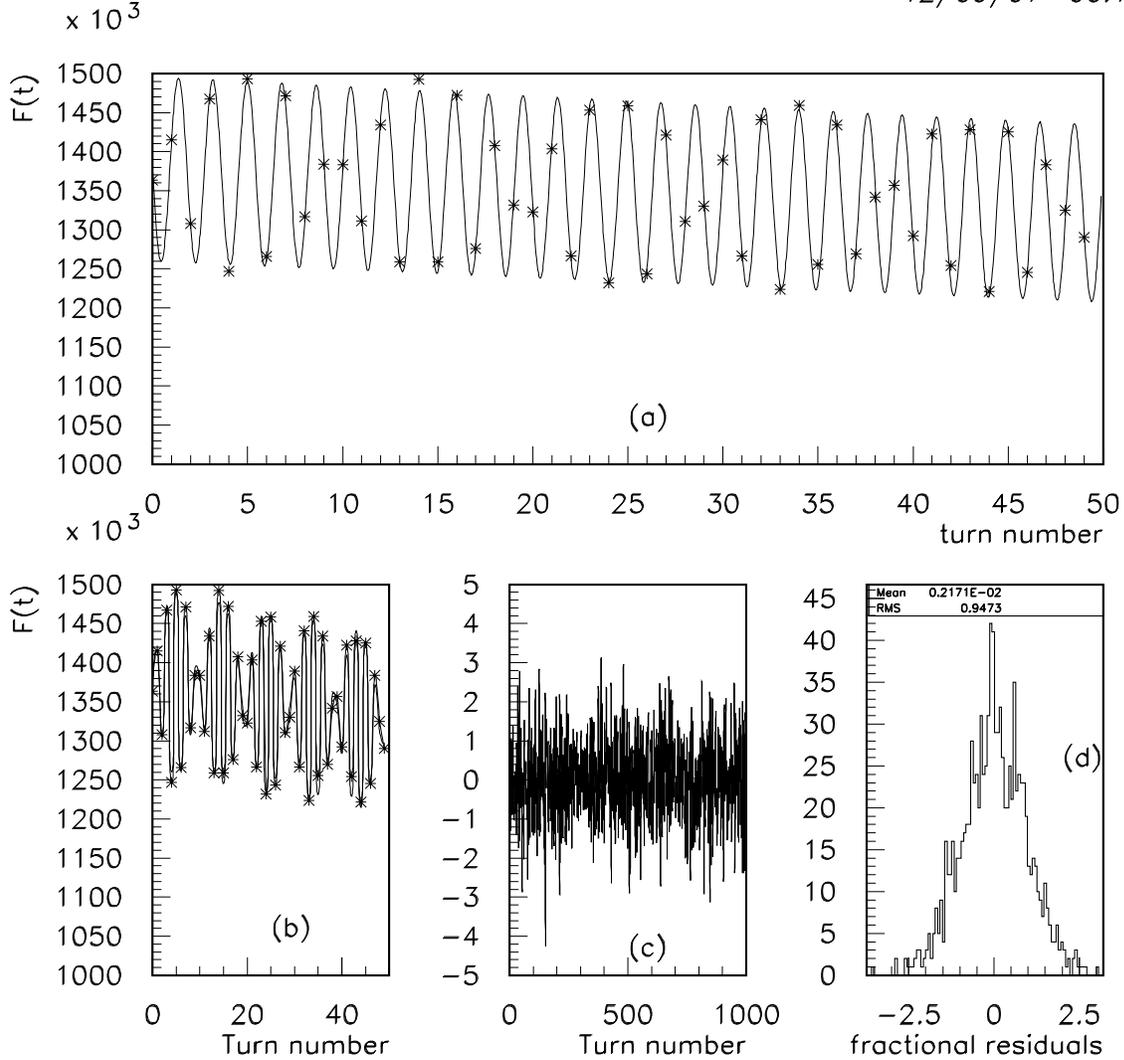}
\caption{(a)Energy detected in the calorimeter during the first 50 turns in a
50 GeV muon storage ring (points). An average value of ${\hat P}$=-0.26 is
assumed and a fractional fluctuation of 0.5E-2 per point. The curve is the
result of a MINUIT fit to the functional form in equation \ref{eq3}. (b) The
same fit, with the function being plotted only at integer turn
values. A beat is evident. (c) Pulls  as a function of turn number (d)Histogram
of pulls.}
\label{fig3}
\end{figure}

\begin{figure}[p]
\epsfxsize = 16.cm
\epsffile{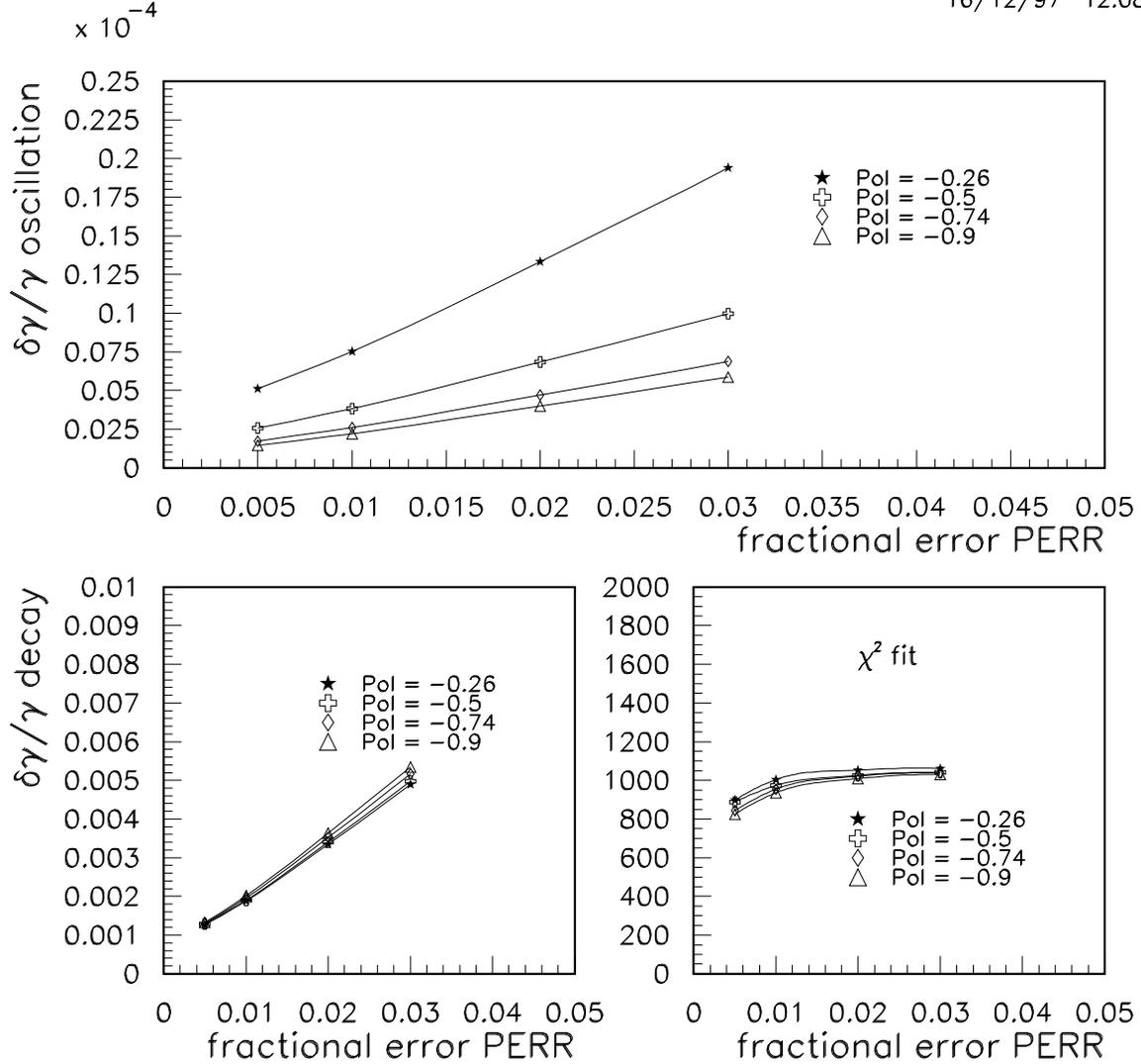}
\caption{(a)Fractional error in $\delta\gamma/\gamma$ obtained from the
oscillations as a function of polarization $\hat P$ and the fractional error in
the measurements PERR (b) Fractional error in $\delta\gamma/\gamma$ obtained
from the decay term as a function of polarization $\hat P$ and the fractional
error in the measurements PERR (c) The total $\chi^2$ of the fits for 1000
degrees of freedom}
\label{fig4}
\end{figure}

We simulate the muon collider spin precession for a grid of values of ${\hat P}$
=-0.9,-0.74,-0.5 and -0.26 and fractional measurement error for the first turn
(PERR) of 0.5E-2, 1.0E-2, 2.0E-2 and 3.0E-2. Figure \ref{fig3}(a) shows the
result of the MINUIT fit plotted for 50 turns for ${\hat P}$=-0.26 and 
PERR=0.5E-2. Figure \ref{fig3}(b) shows the same plot but with the function
being plotted only at integer values of the turn number $t$. A beat is evident
in both the theoretical curve and the simulated measurements as a result of
sampling the oscillation function at fixed intervals, not connected with the
oscillation frequency. The origin of the beat is stroboscopic. Figure
\ref{fig3}(c) shows the pulls, defined as $(data-fit)/error$ at each measurement
as a function of turn number for 1000 turns. There are no major turn dependent
variations in this quantity indicating that the fit converged satisfactorily.
Figure \ref{fig3}(d) shows the histogram of the pulls, which approximates a 
unit Gaussian as desired. Table \ref{tab2} shows the results of the fit for the
grid of values of ${\hat P}$ and PERR.  The results presented in table
\ref{tab2} are shown graphically in Figure \ref{fig4}. As an example, for an
average polarization ${\hat P}$ = -0.26, the fractional error in
$\delta\gamma/\gamma$ varies from 5.1E-6 to 1.9E-5 as the fractional error in
the electron energy sampled varies from 0.5E-2 to 3.0E-2, corresponding to the
number of electrons sampled during the first turn varying from 41261 to 1146.
The average number of decays in the muon collider  is expected to be 3.2E6
decays per meter for a beam intensity of  10$^{12}$ muons. The error in
determining $\gamma$ is thus going to be dominated by the fluctuations in the
number of electrons sampled turn by turn, rather than sampling fluctuations in
the calorimeter. We have simulated conditions involving $\approx$ 40,000 decays.
It should be possible to go to higher statistical  precision than computed here
by sampling larger number of electrons.

 The results for $\delta\gamma/\gamma$ obtained from the measurement of the turn
 by turn  rate  of decay of the electron energy are not competetive with the
 precession method primarily because of the small value of $\alpha$ (0.8399E-3).
 This leads to larger fractional errors for $\gamma$ from this method 
 (which also assumes that the loss of intensity is entirely due to the decay
 process)  by almost  three orders of magnitude than from the precession method.

\subsection{Variation of $\delta\gamma/\gamma$ as a function of muon energy}
The spin precession per turn equals $2\pi$ for a $\gamma$ value of 857.689,
which corresponds to a muon beam momentum of 90.622 GeV/c. This is the first
spin resonance for muons. At this point, the fitting method loses sensitivity
completely, since there will  be no spin oscillations turn by turn. We now study
the error  $\delta\gamma/\gamma$ as a function of beam energy for $\hat P$=-0.26
and PERR=0.5E-2 (keeping the magnetic field in the idealized storage ring to be
4.0 Tesla) as a function of muon beam energy that straddles the spin resonance. 
For initial muon collider physics, the interesting beam energies are 45.5 GeV
(half the Z mass), 80.3 GeV 
\begin{table}[p]
\begin{center}
{\footnotesize
\begin{tabular}{|c|c|c|c|c|c|}
\hline
${\hat P}$ & PERR & Number of electrons & $\delta\gamma/\gamma oscillations$ 
& $\delta\gamma/\gamma decay$  & $\chi^2$ for NDF=1000 \\
           &      & sampled             &            
&                              &                       \\
 \hline
 \hline
      -0.90 & 0.50E-02 &  41261 & 0.14568E-05 & 0.13227E-02 &     824.\\
      -0.90 & 0.10E-01 &  10315 & 0.22147E-05 & 0.20124E-02 &     936.\\
      -0.90 & 0.20E-01 &   2579 & 0.39999E-05 & 0.36398E-02 &    1009.\\
      -0.90 & 0.30E-01 &   1146 & 0.58659E-05 & 0.53457E-02 &    1030.\\
 \hline
      -0.74 & 0.50E-02 &  41261 & 0.17418E-05 & 0.13019E-02 &     843.\\
      -0.74 & 0.10E-01 &  10315 & 0.26183E-05 & 0.19591E-02 &     954.\\
      -0.74 & 0.20E-01 &   2579 & 0.46981E-05 & 0.35229E-02 &    1021.\\
      -0.74 & 0.30E-01 &   1146 & 0.68765E-05 & 0.51672E-02 &    1039.\\
 \hline
      -0.50 & 0.50E-02 &  41261 & 0.25903E-05 & 0.12813E-02 &     888.\\
      -0.50 & 0.10E-01 &  10315 & 0.38407E-05 & 0.19029E-02 &     973.\\
      -0.50 & 0.20E-01 &   2579 & 0.68338E-05 & 0.33972E-02 &    1026.\\
      -0.50 & 0.30E-01 &   1146 & 0.99744E-05 & 0.49749E-02 &    1041.\\
 \hline
      -0.26 & 0.50E-02 &  41261 & 0.51242E-05 & 0.12688E-02 &     898.\\
      -0.26 & 0.10E-01 &  10315 & 0.75317E-05 & 0.18791E-02 &    1004.\\
      -0.26 & 0.20E-01 &   2579 & 0.13324E-04 & 0.33447E-02 &    1053.\\
      -0.26 & 0.30E-01 &   1146 & 0.19380E-04 & 0.48950E-02 &    1061.\\
 \hline
 \hline
\end{tabular}
}
\end{center}
\caption{ Results of fits for  $\delta\gamma/\gamma$ as a function of 
polarization ${\hat P}$ and noise PERR. Also shown is the $\chi^2$ of the fit
for 1000 turns.}
\label{tab2}
\end{table}
($W$ threshold), 175 GeV (top threshold) as well as
half the neutral Higgs mass, which could be as low  as 55 GeV in  some SUSY
scenarios.  We sample all electrons that have energies greater than half the
muon energy. Figure \ref{fig5} shows the variation of $\delta\gamma/\gamma$ as a
function of muon beam energies that straddle these values. It can be seen that
$\delta\gamma/\gamma$ first decreases as one gets close to the resonance and
then blows up on  the spin resonance. Figures (\ref{fig6_2}-\ref{fig6_6}) show
the fitted solutions superimposed on the simulated data for various momenta.
Also shown side by side is the simulated data by itself. As one approaches the
spin resonance, the oscillations slow down. It is nevertheless possible to fit
the slowed down oscillations by a rapidly oscillating theoretical function to
high accuracy on either side of the resonance. At the resonance, the
oscillations die completely, which results in a large value of
$\delta\gamma/\gamma$. It may be possible to use this blow-up in
$\delta\gamma/\gamma$ to find the spin resonance accurately and (paradoxically)
determine $\gamma$ at resonance accurately. This would depend on the width of
the spin resonance, an analysis of which would take us beyond the scope of this
paper.
\subsection{Variation of $\delta\gamma/\gamma$ as a function of beam 
energy spread}
\begin{figure}[p]
\epsfxsize = 16.cm
\epsffile{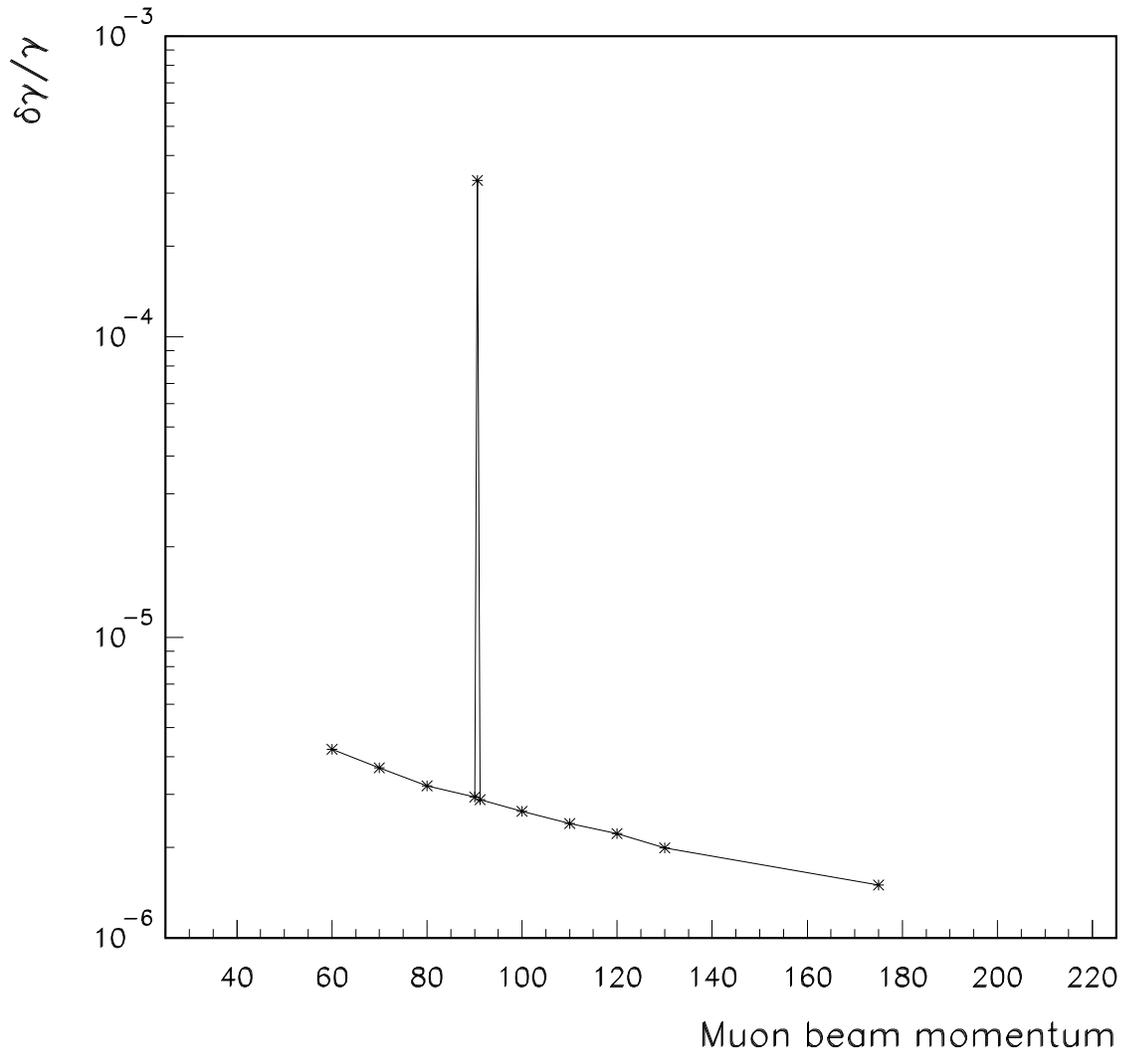}
\caption{Fractional error in $\delta\gamma/\gamma$ obtained from the
oscillations as a function of muon beam momentum}
\label{fig5}
\end{figure}

\begin{figure}[p]
\epsfxsize = 16.cm
\epsffile{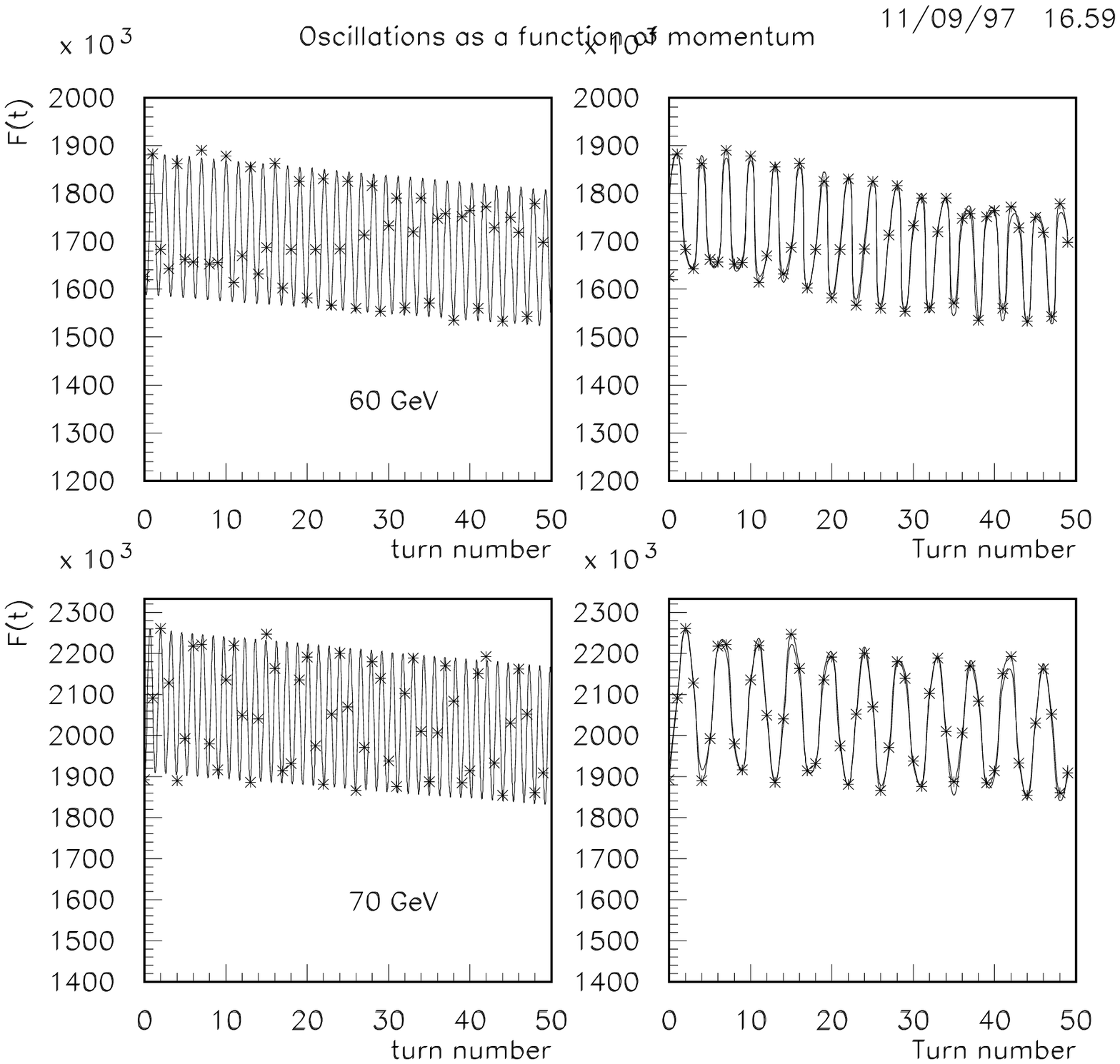}
\caption{The figures on the left hand side show the simulated data with the
fitted function superimposed for 50 turns. The figures on the right hand side
show the simulated data and the fitted function at integer values 
of the turn number. The data shown are 60 GeV/c and 70 GeV/c
muon momenta respectively.}
\label{fig6_2}
\end{figure}

\begin{figure}[p]
\epsfxsize = 16.cm
\epsffile{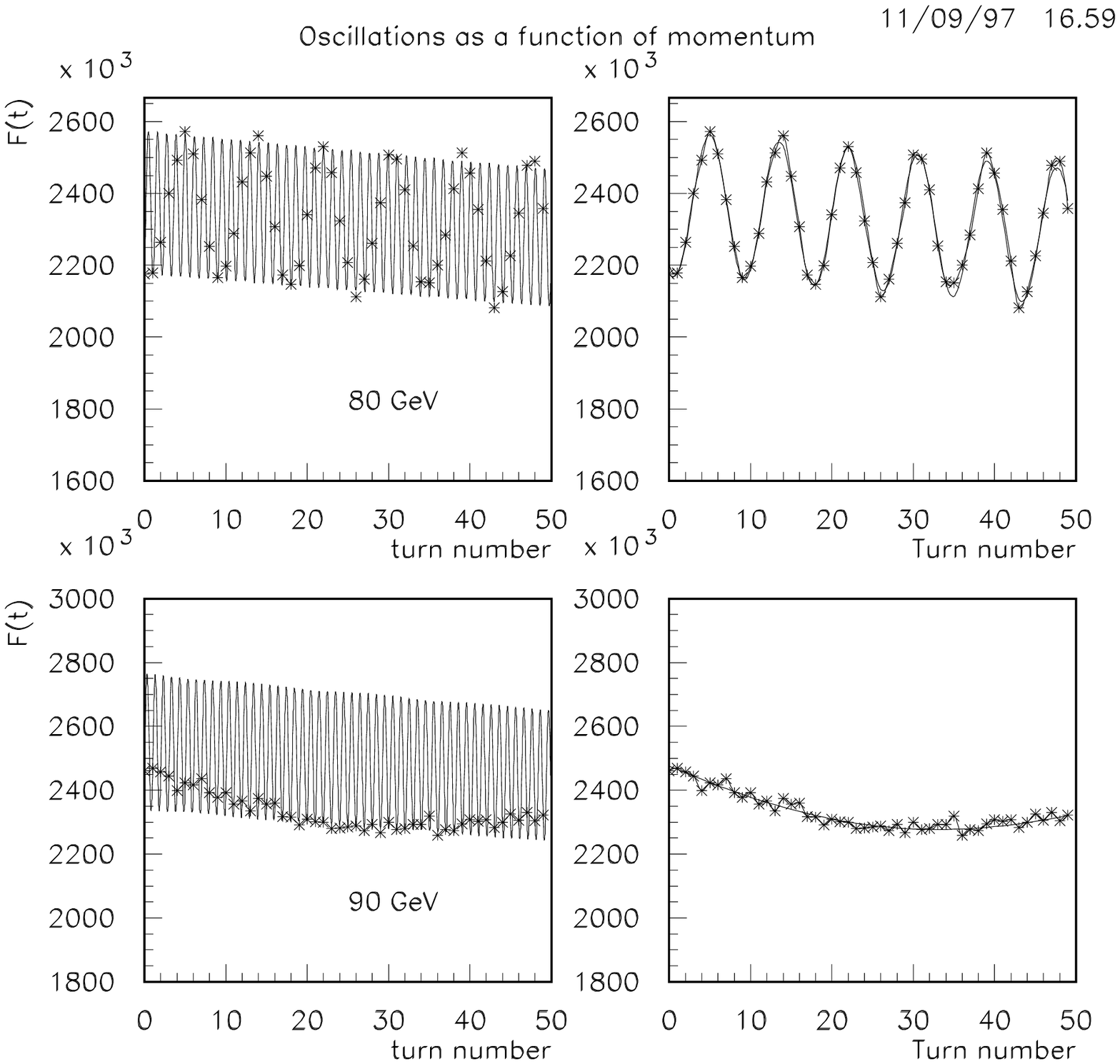}
\caption{The figures on the left hand side show the simulated data with the
fitted function superimposed for 50 turns. The figures on the right hand side
show the simulated data and the fitted function at integer values 
of the turn number. The data shown are 80 GeV/c and 90 GeV/c
muon momenta respectively.}
\label{fig6_3}
\end{figure}

\begin{figure}[p]
\epsfxsize = 16.cm
\epsffile{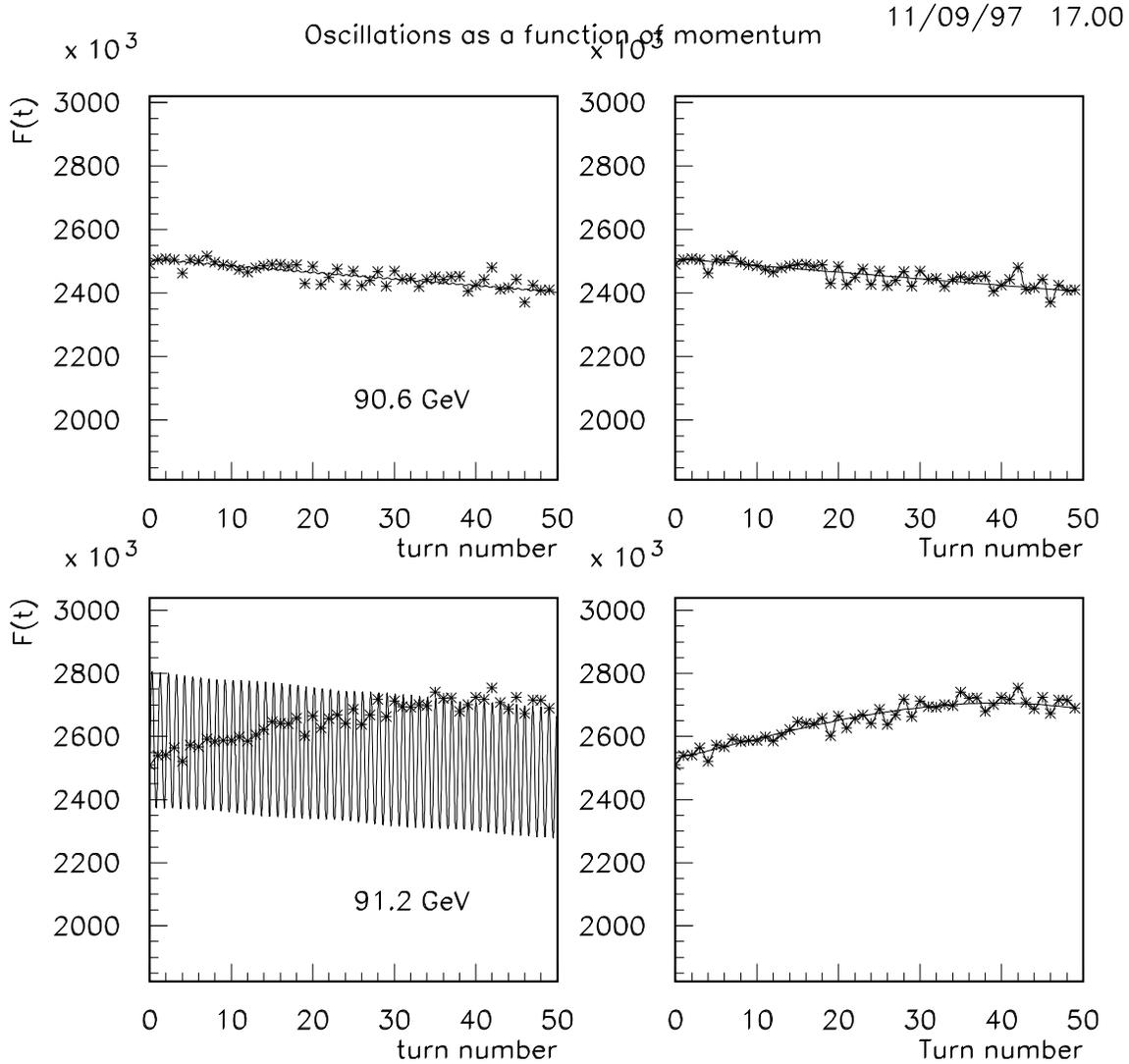}
\caption{The figures on the left hand side show the simulated data with the
fitted function superimposed for 50 turns. The figures on the right hand side
show the simulated data and the fitted function at integer values 
of the turn number. 
The data shown are 90.622 GeV/c and 91.2 GeV/c
muon momenta respectively. The upper curve is on resonance.}
\label{fig6_4}
\end{figure}

\begin{figure}[p]
\epsfxsize = 16.cm
\epsffile{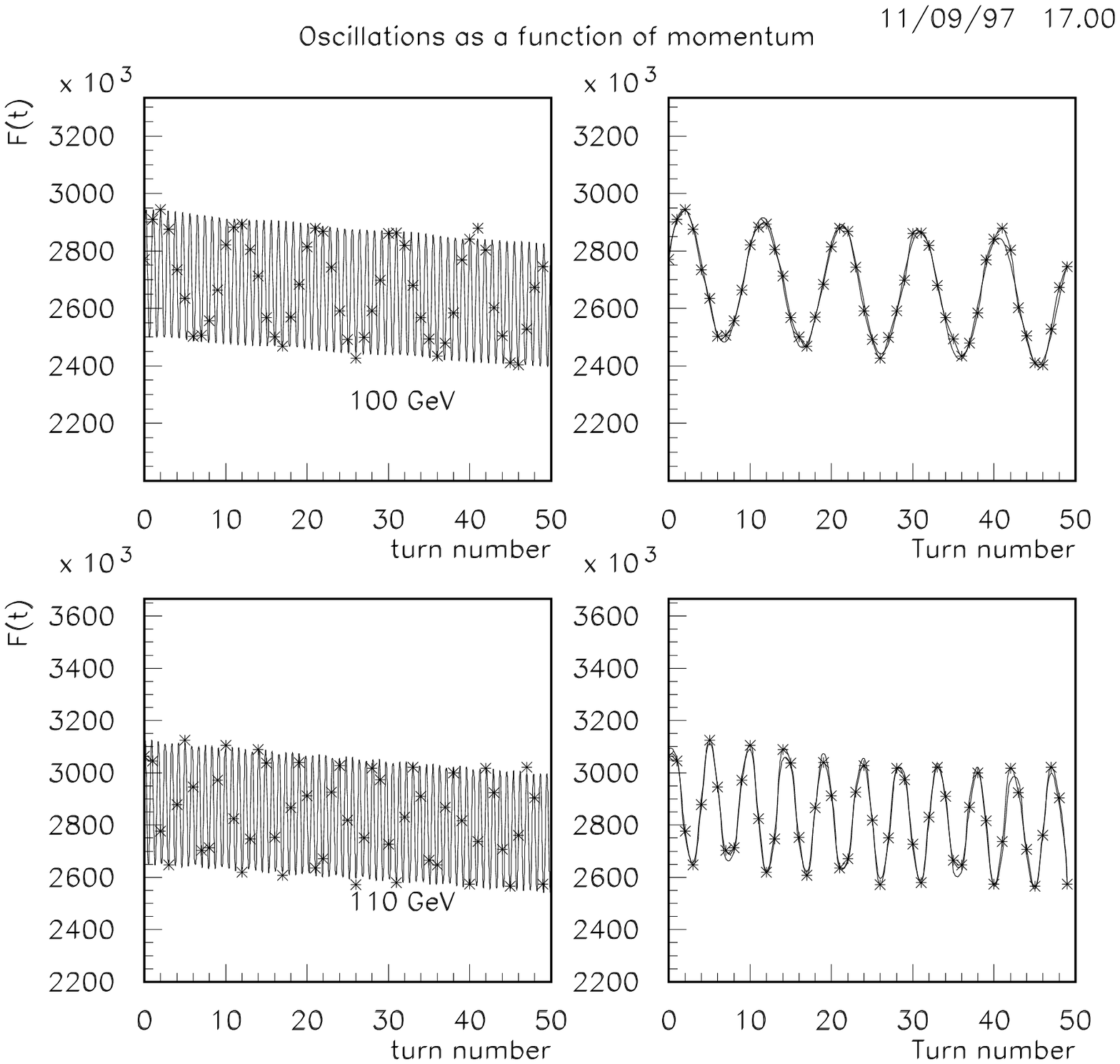}
\caption{The figures on the left hand side show the simulated data with the
fitted function superimposed for 50 turns. The figures on the right hand side
show the simulated data and the fitted function at integer values 
of the turn number. The data shown are 100 GeV/c and 110 GeV/c
muon momenta respectively.}
\label{fig6_5}
\end{figure}

\begin{figure}[p]
\epsfxsize = 16.cm
\epsffile{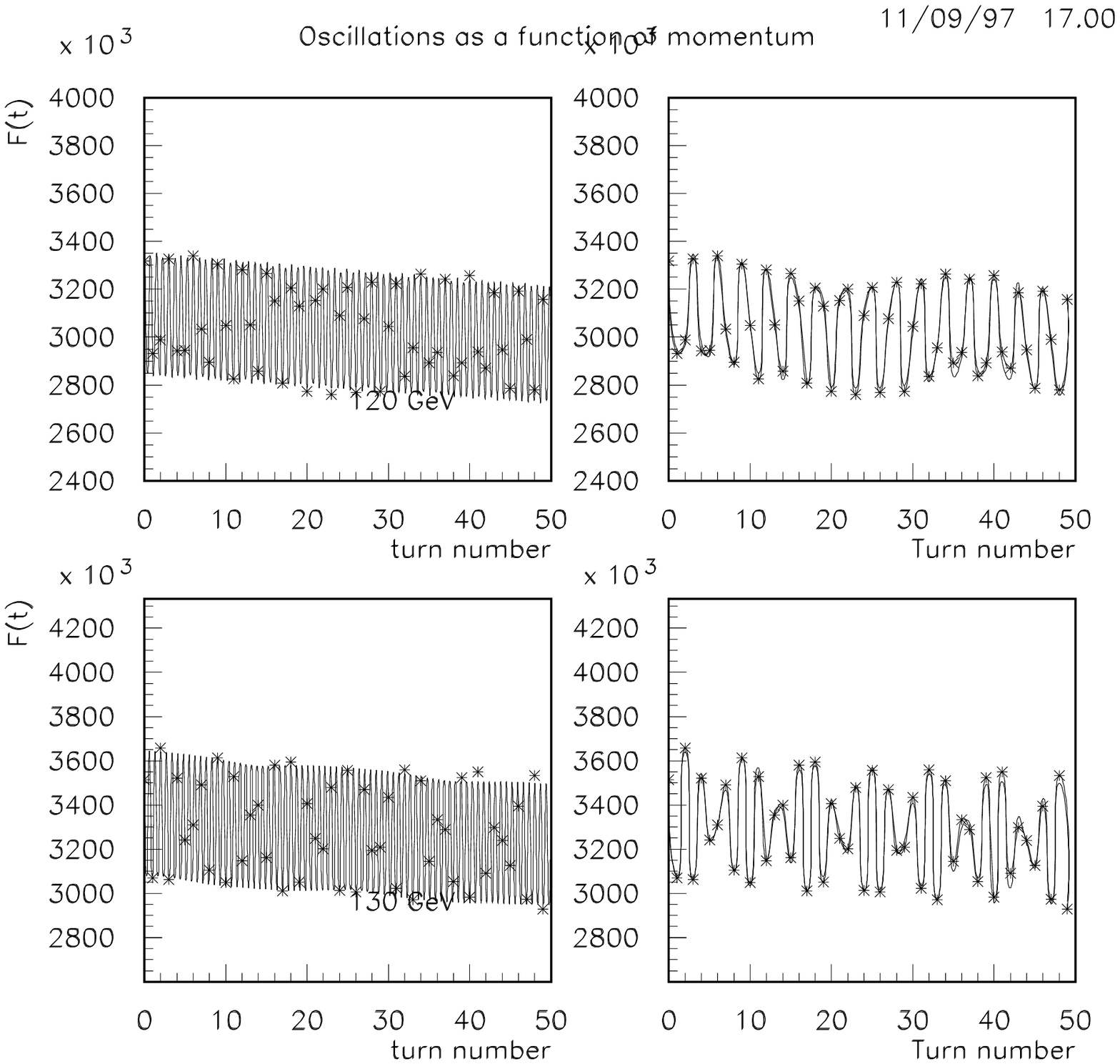}
\caption{The figures on the left hand side show the simulated data with the
fitted function superimposed for 50 turns. The figures on the right hand side
show the simulated data and the fitted function at integer values 
of the turn number. The data shown are 120 GeV/c and 130 GeV/c
muon momenta respectively.}
\label{fig6_6}
\end{figure}

We now calculate the variation of polarization as a function of turn number for
an ensemble of muons with initial value of polarization $\hat P$ = -0.26
and values of momentum spread $\delta p/p $ varying from 0.02E-2 to 0.00125E-2.
This variation is plotted in figure \ref{fig7_0}. For the larger values of
momentum spread, there is a significant degradation of polarization as a
function of turn number, due to differential spin precession of the individual
beam particles. We note that when the beam energy is at 175 GeV, the spin tune
is significantly higher and the depolarization is more rapid. Despite this
depolarization, there is enough information from the first few hundred turns to 
extract the excellent value of $\delta\gamma/\gamma$ for 175 GeV beam energy 
as shown in figure \ref{fig5}.

Figure \ref{fig7} shows the variation of the fractional energy resolution,
$\delta\gamma/\gamma$ as a function of fractional beam energy spread for a muon
beam with $\hat P$ = -0.26, with 41261 electrons sampled. There is
little dependence of $\delta \gamma/\gamma$ on the momentum spread. This is due
to the fact that the momentum spread is determined from the spin tune and not
from the spin oscillation amplitude and the fact that the depolarization is 
not significant for the first few hundred turns for any of the beam  momentum
spreads considered here.
\subsection{Optimization of the electron energy cut}
 We now vary the cut on electron energy and study the dependence on $\delta
 \gamma/\gamma$ on the cut. Figure \ref{fig8} shows the variation of
 $\delta\gamma/\gamma$ with the cut on individual electron energies for
 $\hat P$ = -0.26 for 41261 and 1146 electrons sampled. As shown in the
 Appendix, the fractional error on the average energy of electrons is much
 smaller than the fractional error on the total energy of electrons. It is
 possible to measure the average electron energy by counting the number of
 electrons going into the calorimeter with a scintillator array. However, the
 precession information is contained increasingly in the number of electrons
 rather than their average energy as we increase the electron energy cut.
Figure \ref{fig8} shows the variation of $\delta\gamma/\gamma$ calculated from
average as well as total electron energy as a function of the electron 
energy cut. For smaller values of the electron energy cut, the average method
produces superior errors than the total energy method. However, with 40,000
electrons or more sampled a total energy method with a cut of 25 GeV or higher
seems optimal. It should however be pointed out that the average energy method
does not require a model for the rate of decay of muon intensity in the machine,
which in practice could be a complicated function of turn number. As such the
systematics associated with this would not be present in the average energy
method.
Figure \ref{fig9}(a) shows the variation of the absolute value of $C/F$ as a
function of the electron energy cutoff for $\hat P$ = -0.26, where $C$ and $F$
are defined in equation  \ref{eq3} for both the total energy method and the
average energy method. Figure \ref{fig9}(b) shows the fraction of
electrons that lie above the electron energy cut as a function of the energy
cut. The polarization for this sample is 0, since the electron energy fraction
depends on polarization as well. Given the curves shown in figure \ref{fig9}, it
should be possible to estimate the error in $\delta\gamma/\gamma$ for a variety
of conditions.
\section{Effects due to departures from the ideal case}
 So far we have considered a planar collider ring with uniform vertical 
magnetic field  and no electric fields. 
The actual collider ring will depart from the ideal in three respects; a)It
will have RF electric fields to keep the muons bunched, b) it will have radial 
horizontal magnetic fields experienced by partcles in an  off-center trajectory
at quadrupoles and at vertical correction dipoles, and c) it will have
longitudinal magnetic fields due to solenoidal magnets in the interaction
region(s).  We now consider the effect due to  each of these departures from
the ideal.
\subsection{Electric fields}
 Equation \ref{bmt} implies that there is no spin precession due to 
longitudinal electric fields ($\vec{\beta} \times \vec{E}$ = 0). RF electric 
fields are longitudinal, so there will be no precession due to the RF electric
fields. At present there are no plans to install electrostatic separators to
separate the beams. If and when this happens, one should consider the effect
due to the transverse electric fields thus introduced.
\subsection{Effect of radial magnetic fields}
 Particles which are off-axis at quadrupoles will experience radial as well as 
vertical magnetic fields. Even though the net integral of these off-axis fields
around the ring is zero, the spin rotation along a horizontal axis followed by
spin rotation about a vertical axis (caused by a bend dipole) followed by a
reverse rotation in the horizontal direction still produces a net effect since
the rotations about the horizontal and vertical axes do not commute. The
effects have been analyzed by Assmann and Koutchouk \cite{ass} who show that 
this results in both a net spin tune shift $<\delta\nu>$ 
as well as a spread in tune $\sigma_{\delta\nu}$.
\begin{equation}
 <\delta \nu> =  \frac{cot\pi\nu_0}{8\pi} \nu_0^2 \left(n_Q (Kl_Q)^2 \sigma^2_y
+ n_{CV} \sigma^2_{\theta CV} \right)
\label{eq4}
\end{equation}
where $\nu_0 \equiv a\gamma$ is the spin tune of the collider ring, 
$n_Q$ are the number of quadrupoles with integrated gradient $Kl_Q$,
$\sigma_y$ is the misalignment spread of the closed orbit at the quadrupoles, 
$n_{CV}$ is the number of vertical correction dipoles
and $\sigma_{\theta CV}$ is the rms beand angle in the vertical correctors.
The spread in tune is given by,
\begin{equation}
 \sigma_{\delta \nu} =  \frac{<\delta\nu>}{cos\pi\nu_0}
\end{equation}
Table \ref{tab3} shows the values for $<\delta\nu>$ and $\sigma_{\delta \nu} $
obtained by Assman and Koutchuk \cite{ass} for LEP. We compare this with 
to the current design for the 50 GeV muon collider ring \cite{johnstone}.
Including the low beta section, there
are 70 quadrupoles with an RMS value of $Kl_Q$ = 0.27 m$^{-1}$. The effects
due to correction dipoles may be neglected in both the LEP and the muon collider
cases. We assume a beam misalignment of 5mm at the quadrupoles, which is the
same value used in the LEP calculation. This is probably being conservative. The
tune shift for LEP corresponds to a shift in beam energy calibration of 3.0 KeV.
The tune spread for LEP corresponds to a spread in beam energy calibration of 30
KeV. For the muon collider, the tune shift corresponds to a shift in beam energy
calibration of -0.24 KeV and a spread of 1.46 KeV, both of which are negligible.
The reason for the smallness of this effect for the muon collider is twofold.
Since the circumference of the muon collider is smaller than LEP, there are
fewer quadrupoles.
Secondly, the muon is two hundred times more massive than the electron and has
has a spin  tune $a\gamma$ that is smaller by the same factor.  The spin tune
shift depends on the  the square of the spin tune. It should be noted that the
above formulae are not valid for a fractional spin tune of 0.5. 
\begin{table}[p]
\begin{center}
{\footnotesize
\begin{tabular}{|c|c|c|c|c|c|c|}
\hline
 Machine & Spin tune $\nu_0$ & Quadrupoles & RMS $Kl_Q$  & 
$\sigma_y$ & $\delta\nu$ &  $\sigma_{\delta\nu}$ \\
         &                   &             &            meters$^{-1}$     &
 meters    &             &                       \\
\hline
 46 GeV LEP     & 100.47 & $\approx$ 600 & 0.032 & 0.5E-3 & 5.7E-6 & 6.1E-5 \\
                &        &    &       &     & $\equiv$ 3KeV & $\equiv$ 30KeV \\
\hline
 50 GeV Muon Collider & 0.5517 & 70 & 0.274 & 0.5E-3 & -0.26E-8 & 1.66E-8 \\
                &        &    &    &   & $\equiv$ -0.24KeV & $\equiv$ 1.46KeV \\
\hline
\end{tabular}
}
\end{center}
\caption{ Predictions for spin tune shift $\delta\nu$ and 
spread in spin tune shift $\sigma_{\delta\nu}$ caused by quadrupoles 
for LEP compared to the 50 GeV muon collider ring}
\label{tab3}
\end{table}

\subsection{Solenoidal  magnetic fields}

The experimental region will in all likelihood contain a solenoidal magnet.
This solenoidal field, if uncorrected, will rotate the spin vector of the muons
about the beam direction by a constant amount $\theta_s$ 
per turn, which can be derived using equation \ref{bmt}.
\begin{equation}
 \theta_s = - \frac{e}{\gamma m_\mu} (1 + a) B_s = -(1+a)\frac{B_sl}{B\rho}
\end{equation}
where $B_s$ is the field due to the solenoid of length $l$, $B$ is the dipole
bending field of the ring of radius $\rho$. For a solenoid of 1.5 Tesla and
length 6 meters, $\theta_s$ = 3.09 degrees for the planar storage ring
parameters of table \ref{tab1}. It can be shown analytically \cite{kout} that
this produces a spin tune shift $\delta\nu$ given by
\begin{equation}
 \nu + \delta\nu = \frac{1}{\pi} arccos \left( cos (\pi\nu)cos(\frac{\theta}{2})
 \right)
\end{equation}
yielding  a spin tune shift $\delta \nu$ = -1.901E-5, or a fractional spin tune
shift of $\delta\nu/\nu$ = -3.45E-5. For a 50 GeV muon beam, this is a shift in
energy calibration of -1.72 MeV. In LEP, a similar solenoid will have a much
smaller fractional tune shift \cite{kout},  since the tune is 200 times larger
for electrons. It is important to correct the effect due to the solenoids, since
this is cumulative turn by turn.  At LEP this is done by a series of vertical
orbit correctors\cite{ross}  followed by normal lattice followed by vertical
orbit correctors of reverse polarity, which has the effect of rotating the spin
by half the amount produced by the solenoid. A similar set of  corrections is
inserted after the solenoid to complete the correction. This method depends
on a non-zero value of $g-2$ and as such will be 200 times less effective for
muons than for electrons, for any given magnet strength. The most effective
method to correct for the solenoid is to surround it on either side by
compensating solenoids of  minimal radius large enough to allow the beam 
to go through.
\section{Conclusions}
We have demonstrated that it is feasible to measure the energy of a 50 GeV muon
collider to a few parts per million using the $g-2$ spin precession technique,
provided it is feasible to maintain a muon polarization of the order of $\hat
P$=0.25 in the ring for a thousand turns. In order to explore the Higgs
resonance, it is necessary to measure the bunch by bunch variation in energy to
a few parts per million. We have demonstrated that the $g-2$ technique is
capable of doing so. It is still possible to tolerate a spin tune shift in the
overall energy scale of a few percent, which will act only as a systematic error
on the Higgs mass and width.

We would also like to note in passing that polarization information from a
calorimeter of the type proposed  here can be used in conjunction with a
neutrino detector placed along the line of the neutrinos produced in association
with the electrons to estimate the variation in the energy spectrum of the muon
neutrinos and electron antineutrinos in the beam. Such information can be a
valuable tool in untangling various possible neutrino oscillation scenarios.

We intend to develop the method here by studying the propagation of polarized
muons in a realistic 50 GeV collider lattice using the program COSY \cite{cosy}
that takes  into account non-linear effects in the dynamic aperture. Design and
Monte Carlo studies will also be undertaken to develop the calorimeter detector
needed.
The authors would like to  acknowledge useful conversations with Alain Blondel,
Yaroslav Derbenev and Robert Rossmanith.
\begin{figure}[p]
\epsfxsize = 16.cm
\epsfysize = 16.cm
\epsffile{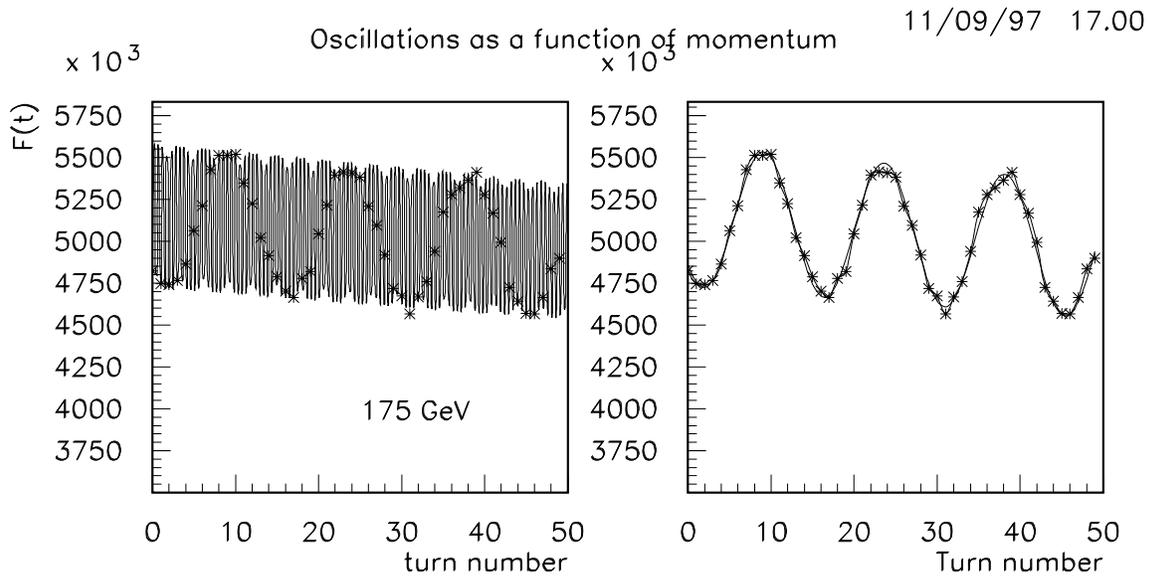}
\caption{The figures on the left hand side show the simulated data with the
fitted function superimposed for 50 turns. The figures on the right hand side
show the simulated data and the fitted function at integer values 
of the turn number. The data shown are 175 GeV/c muon
momentum.}
\label{fig6_7}
\end{figure}
\begin{figure}[p]
\epsfxsize = 16.cm
\epsfysize = 16.cm
\epsffile{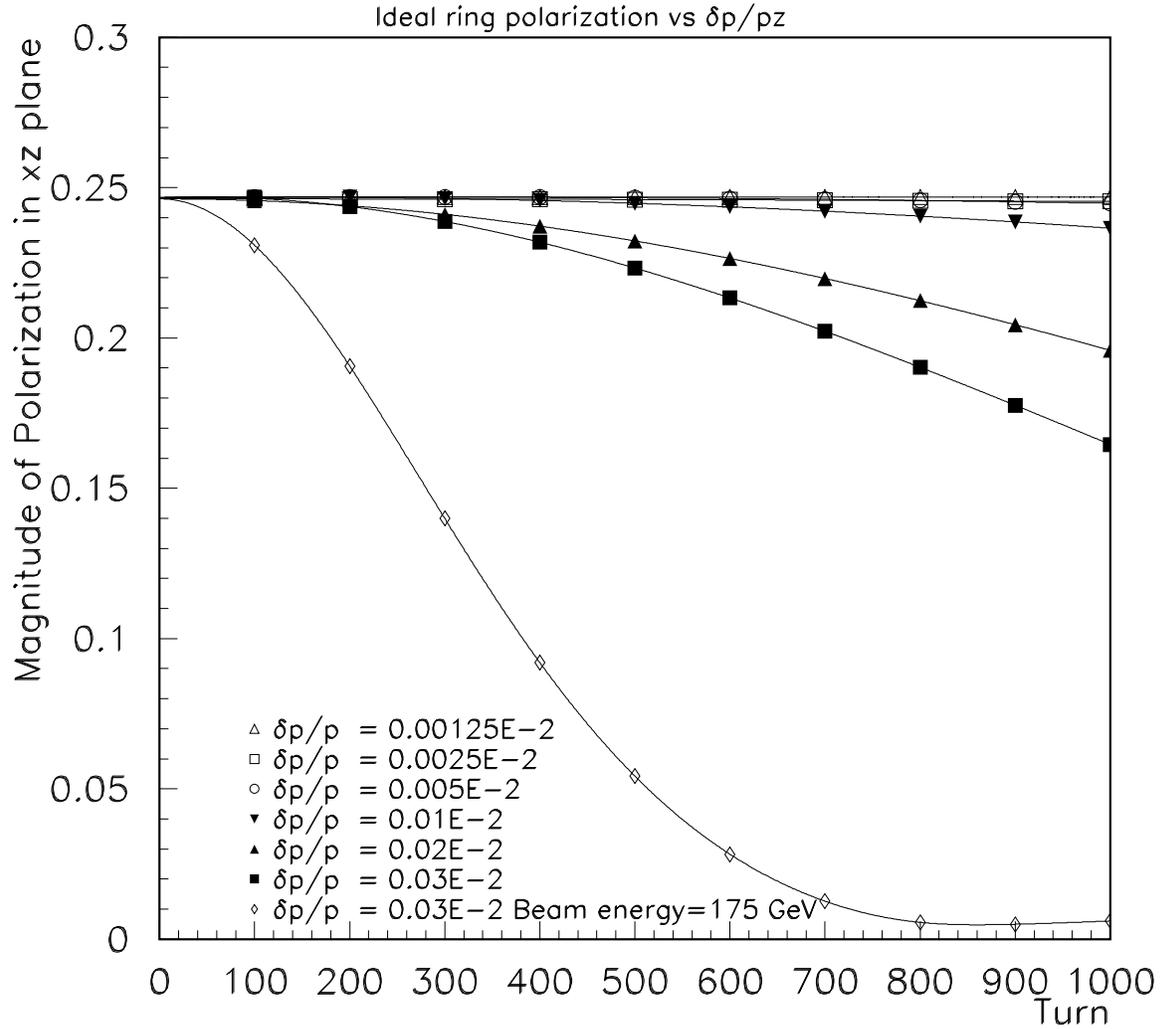}
\caption{ Variation of polarization as a function of turn number for 
50 GeV muons with initial $\hat P$ =-0.26 and various values of $\delta p/p$
in an ideal collider ring. The bottom curve is for 175 GeV muons and shows a
more rapid depolarization due to the higher spin tune.}
\label{fig7_0}
\end{figure}
\begin{figure}[p]
\epsfxsize = 16.cm
\epsfysize = 16.cm
\epsffile{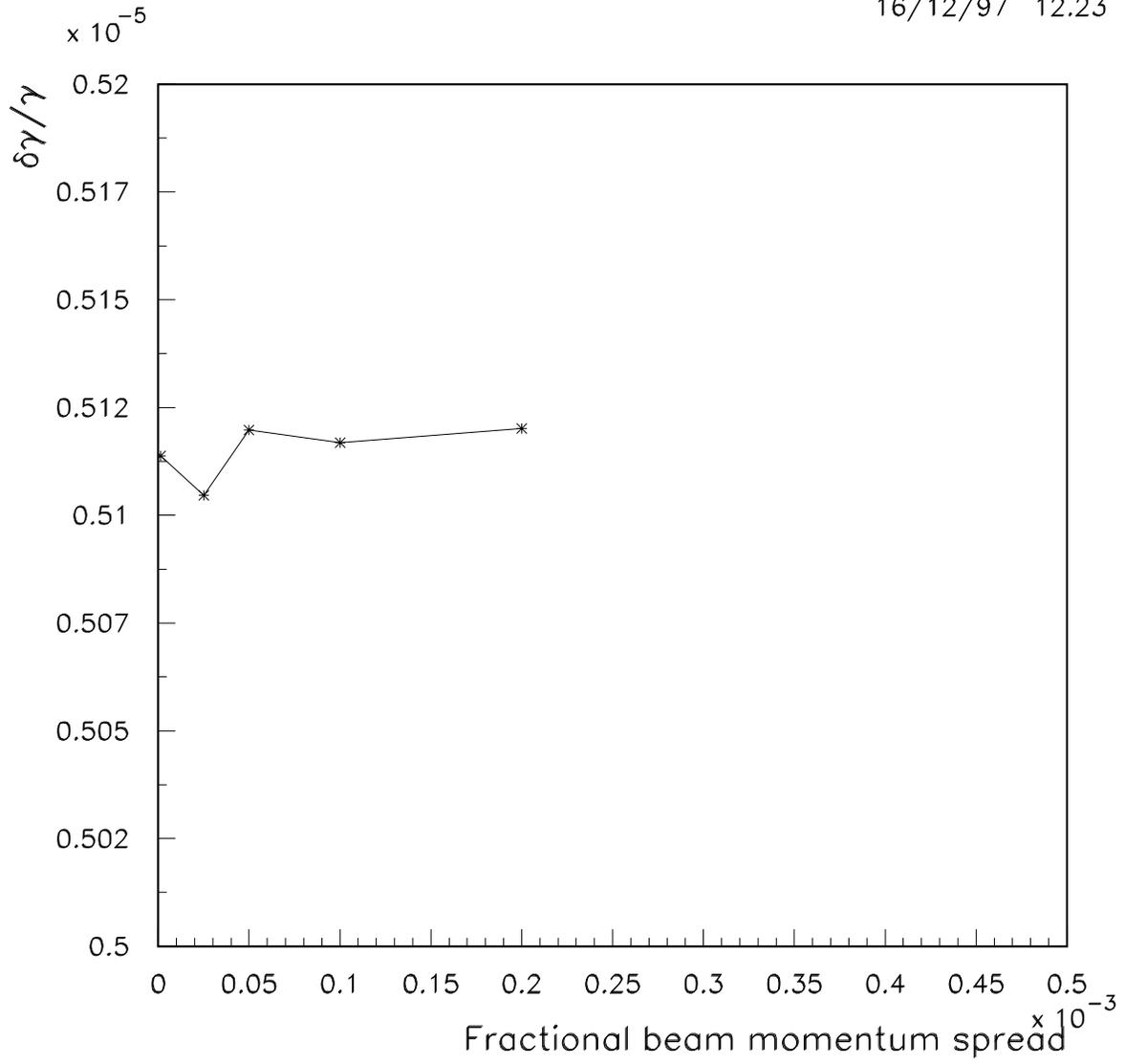}
\caption{ $\delta\gamma/\gamma$ versus fractional beam energy spread for 50 GeV
muons with PERR=.5E-2 and $\hat P$ =-0.26}
\label{fig7}
\end{figure}
\begin{figure}[p]
\epsfxsize = 16.cm
\epsffile{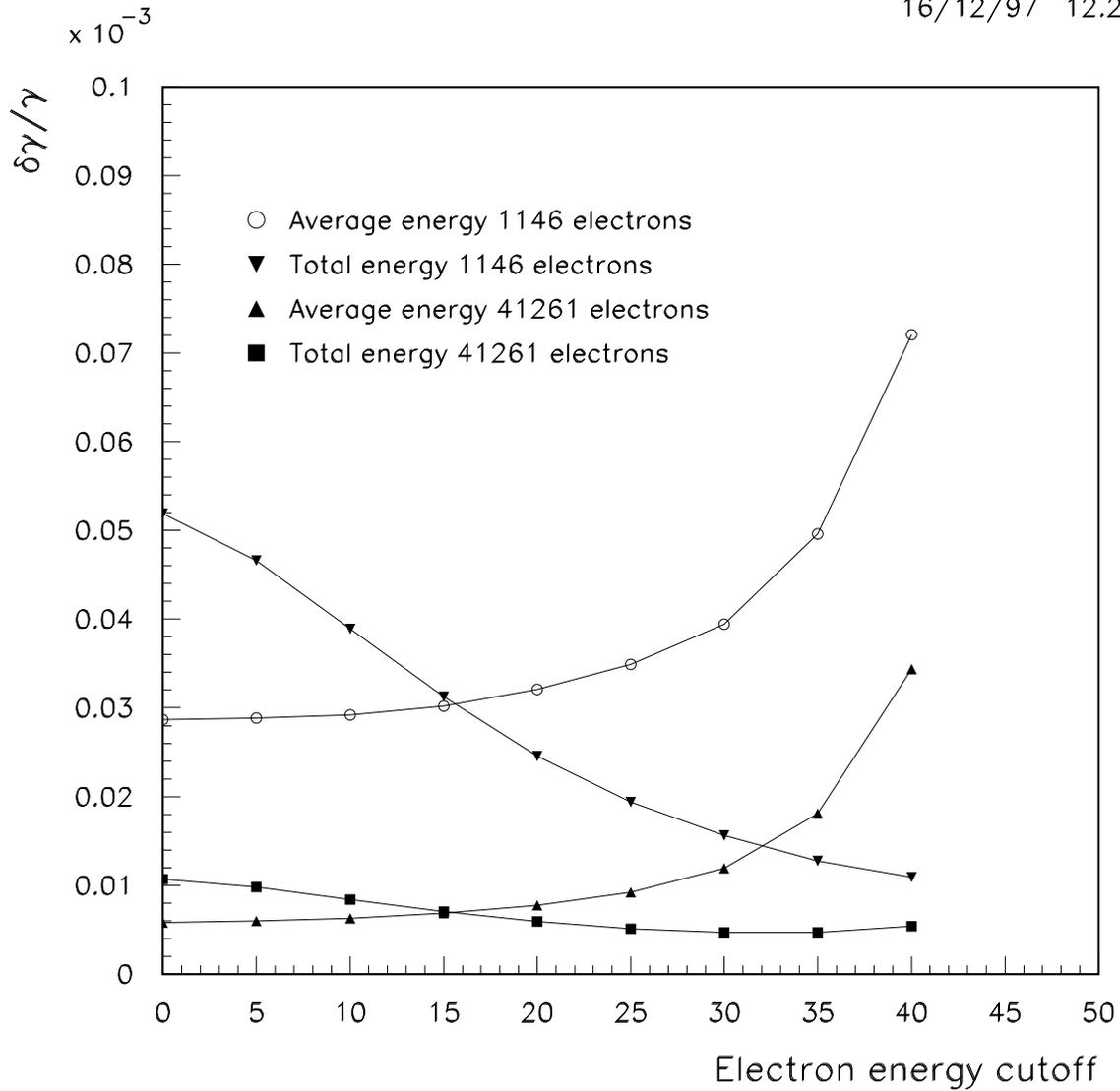}
\caption{ The variation of $\delta\gamma/\gamma$ as a function of the electron
energy cut for 41261 and 1146 electrons  $\hat P$ =-0.26. We 
fit the total energy in the calorimeter as well as the average energy per
electron}
\label{fig8}
\end{figure}
\begin{figure}[p]
\epsfxsize = 16.cm
\epsffile{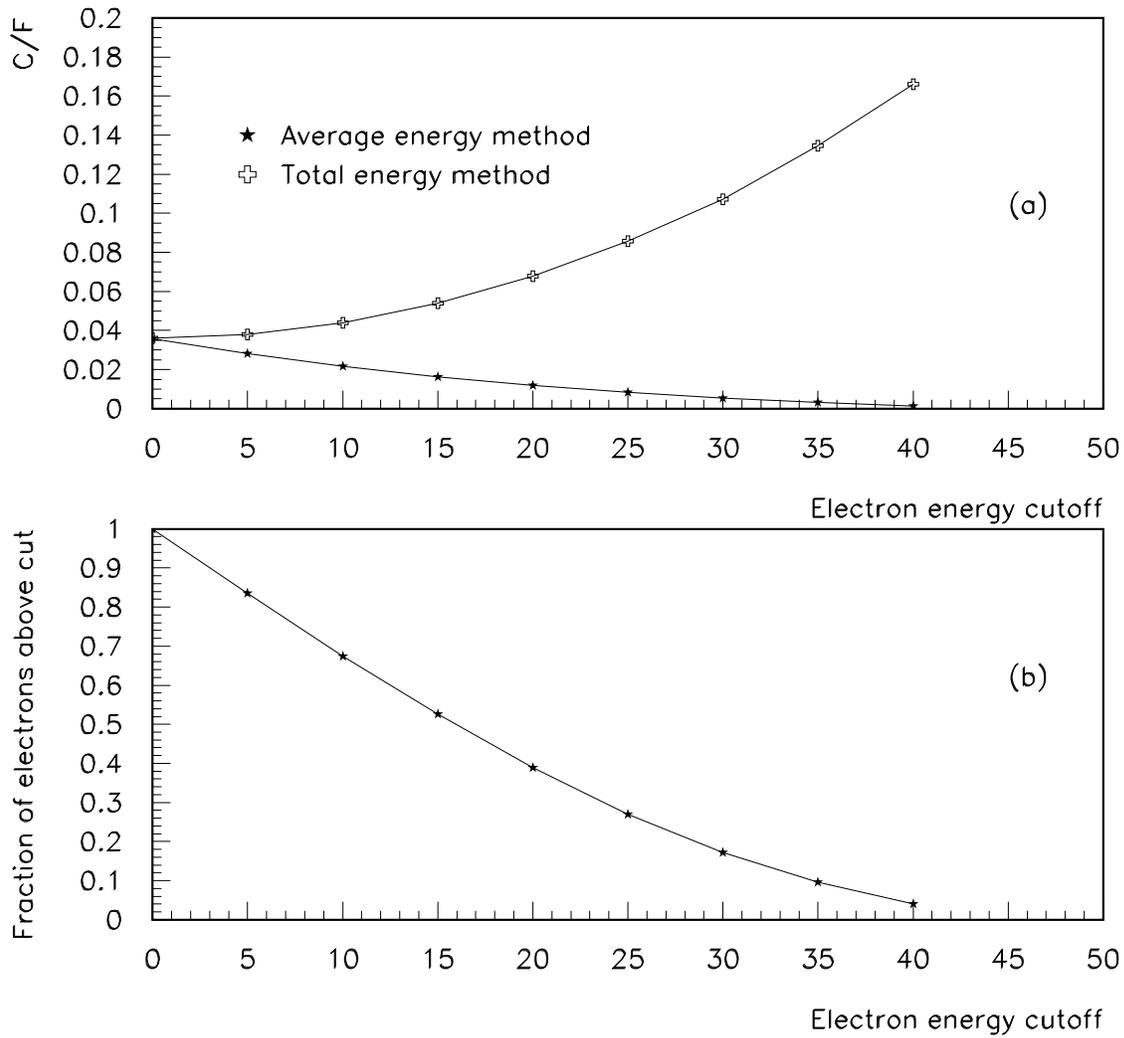}
\caption{(a) The variation of $C/F$ as a function of the electron
energy cut for $\hat P$ =-0.26 for total energy method and average energy 
method. (b) The fraction of electrons that survive the energy cut as a function
of the cut for $\hat P$ = 0.}
\label{fig9}
\end{figure}

\newpage
\section{appendix}
\subsection{Treatment of Errors}

We measure the total energy  $E$  of all electrons with individual energy  $e>$
25 GeV in an electromagnetic calorimeter. Let N be the  number of electrons
sampled during a turn. N can fluctuate from sampling to sampling. Then
\begin{eqnarray}
 E = \sum_{i=1}^{N} e_i = N<e>\\
 \frac{\sigma^2_E}{<E>^2} = \frac{\sigma^2_N}{<N>^2} + 
 \frac{\sigma^2_{<e>}}{<e>^2} 
=\frac{1}{<N>}(1 + \frac{\sigma_e^2}{<e>^2}) \label{eqa1}
\end{eqnarray}
where the variance $\sigma^2$ of the quantities $e$ and $E$ 
results from the kinematic distributions of those quantities and not from the
measurement errors. The average of the individual electron energies is denoted 
by $<e>$.

Let the calorimeter be such that it measures the true deposited energy $E$ with
a resolution $\epsilon(E)$ such that
\begin{equation}
 \frac{\epsilon^2}{E^2} = {\cal C}^2 + \frac{{\cal S}^2}{E} + 
 \frac{{\cal N}^2}{E^2}
\label{eqa2}
\end{equation}
where $\cal{C},\cal{S} $ and $\cal{N}$ represent the Constant, Sampling and
Noise terms respectively. Let us assume that the measurement errors are
Gaussian. Then,
\begin{equation}
   P(E_m) = \int P(E) G(E,E_m,\epsilon) dE
\end{equation}
where $E_m$ is the measured energy and G(E,$E_m,\epsilon$) is a Gaussian of mean
$E$ and standard deviation $\epsilon$, which is a function of $E$ and is written
as $\epsilon(E)$.  From this it follows that the mean measured energy $<E_m>$
and the mean squared measured energy $<E_m^2>$ are given by
\begin{eqnarray}
\begin{array}{l}
 <E_m> = \int E_m  P(E_m)dE_m  = \int E_m dE_m \int P(E) G(E,E_m,\epsilon) dE \\
= \int P(E) dE \int E_m G(E,E_m,\epsilon) dE_m = \int P(E) dE \times E = <E> 
\label{eqa3} \\
\end{array}
\end{eqnarray}
The equation \ref{eqa3} states that the mean value of any distribution given by 
$P(E)$  is the same as that of the smeared distribution $P(E_m)$ provided the
smearing function is such that the average of the smeared values for any given
true value $E$ equals the true value ( a property satisfied by Gaussians) and
the integration is carried over the full range of the variables. As an aside, in
High Energy Physics, we measure steeply falling spectra that are smeared by
measurement errors. Provided there is no arbitrary lower cut-off in the measured
spectra (such as a trigger threshold), the above result would be valid, even for
non-Gaussian resolutions. For the muon collider, the cut off in selected
electrons of 25 GeV is imposed by momentum selection that is independent of the
calorimetry. So the above result would still  be valid. Similarly, one can
compute $<E_m^2>$
\begin{eqnarray}
\begin{array}{l}
 <E_m^2> = \int E_m^2  P(E_m)dE_m  = 
\int E_m^2 dE_m \int P(E) G(E,E_m,\epsilon) dE \\
= \int P(E) dE \int E_m^2 G(E,E_m,\epsilon) dE_m = 
 \int P(E) dE \times (E^2 + \epsilon^2)  = <E^2> + \int P(E) \epsilon^2(E) dE 
\end{array}
\end{eqnarray}

From this it follows that the variance of the measured energy $\sigma^2_{E_m}$
is given by
\begin{equation}
 \sigma^2_{E_m} = \sigma^2_E + \int P(E) \epsilon^2(E) dE \\
               \approx \sigma^2_E  + \epsilon^2(<E>)
\end{equation}
where the last approximation results from assigning the  average measurement
resolution to the resolution at the average energy. This then leads to
\begin{equation}
\frac{\sigma^2_{E_m}}{<E_m>^2} \approx \frac{\sigma^2_E}{<E>^2} + 
\frac{\epsilon^2(<E>)}{<E>^2}
\end{equation}
Using equation \ref{eqa1} and \ref{eqa2} leads to
\begin{equation}
\frac{\sigma^2_{E_m}}{<E_m>^2} \approx \frac{1}{N}
(1 + \frac{\sigma^2_e}{<e>^2}) +
{\cal C}^2 + \frac{{\cal S}^2}{N <e>} + 
 \frac{{\cal N}^2}{N^2 <e>^2}
\end{equation}
From the above equation, it is obvious that the calorimeter must be such that
the constant term ${\cal C}$ must be negligible for the fractional 
resolution to scale inversely with the number $N$ of electrons collected. The
noise term can be neglected for large enough $N$ since it goes as $N^{-2}$.
With these assumptions, one gets
\begin{equation}
\frac{\sigma^2_{E_m}}{<E_m>^2} \approx 
 \frac{1}{N}(1 + \frac{\sigma^2_e}{<e>^2} +
  \frac{{\cal S}^2}{<e>})
\end{equation}
For a  50 GeV muon beam, the values of $<e>$ and $\sigma_e$ are 34.05 GeV and 
6.046 GeV respectively for electrons with $e>$ 25 GeV. The ratio, $\sigma_e/<e>$
is to a good approximation independent of muon energy. This then leads to the
following error formula.
\begin{equation}
\frac{\sigma^2_{E_m}}{<E_m>^2} \approx \frac{1}{N}(1 + 0.03153 + 
  \frac{{\cal S}^2}{34.05})
\end{equation}
Sampling terms of 0.15 GeV$^{1/2}$ or better are easy to obtain in
electromagnetic calorimeters. This leads to 
\begin{equation}
\frac{\sigma^2_{E_m}}{<E_m>^2} \approx \frac{1}{N}(1 + 0.03153 +
  0.000661)
\end{equation}
i.e. the sampling term can be neglected when compared to the fluctuation in the
true electron energies.
So if the fractional measurement error is PERR $\equiv
\frac{\sigma_{E_m}}{<E_m>}$ is specified, the equivalent number of electrons
is given by
\begin{equation}
 N \approx \frac{1.03153} {(PERR^2)}
\end{equation}
In other words, PERR=.5E-2,1.0E-2,2.0E-2 and 3.0E-2 implies 41261, 10315, 2579,
and 1146 electrons sampled. If in practice we sample 100,000 electrons,  this
leads to a value of PERR=0.3212E-2. In order for this good a resolution to be
meaningful, the constant term ${\cal C}$ has to be below this order of
magnitude.

\subsection{Using averages}
Equation \ref{eqa1} holds for the total energy $E$ in the calorimeter. If
however, one also measures the total number of particles entering the
calorimeter (using a scintillator system for example, that counts minimum
ionizing particles), then for each turn one can measure the average energy
$E_{av}$ of electrons. The fractional error on $E_{av}$ does not contain a term
due to the fluctuation of the number of electrons entering the calorimeter,
being given by 
\begin{equation}
 \frac{\sigma^2_{E_{av}}}{<E_{av}>^2} =  \frac{\sigma^2_{<e>}}{<e>^2} 
=\frac{1}{<N>}(\frac{\sigma_e^2}{<e>^2}) \label{eqa4}
\end{equation}
with $<E_{av}> = <e>$. For a fractional error of PERR in $E_{av}$, the
equivalent number of electrons sampled would be given by
\begin{equation}
 N \approx \frac{0.03153} {(PERR^2)}
\end{equation}

With this method,
PERR=.5E-2,1.0E-2,2.0E-2 and 3.0E-2 implies 1261, 315, 79,
and 35 electrons sampled, assuming no error in the measurement of N.
If we sample 100,000 electrons, the fractional error in the average would be
0.561E-3. For this error to be meaningful, the sampling term would have to
be of this order of magnitude.

\end{document}